\documentclass[aps,preprint]{revtex4}

\newcommand{\be}{\begin{equation}}

\newcommand{\ee}{\end{equation}}
\newcommand{\bea}{\begin{eqnarray}}
\newcommand{\eea}{\end{eqnarray}}

\usepackage{graphicx}  % got figures? uncomment this
\usepackage{amsmath}
\usepackage{bm}

\begin{document}

\title{On the critical end point of the QCD and the NJL model phase diagrams}
\author{M.~Ruggieri}\affiliation{Universit\`a degli Studi di Bari, Italy}
  \affiliation{Istituto Nazionale di Fisica Nucleare, Sezione di Bari, Italy}
%\PACSes{\PACSit{00.00}{**************************************}}

\begin{abstract}
In this talk I compare the knowledge on the critical end point of the QCD phase diagram grasped from lattice
calculations, with that obtained from Nambu--Jona-Lasinio~(NJL) model computations. The original publication is
available at ``http://www.sif.it/SIF/en/portal/journals''.
\end{abstract}

\maketitle

%\section{Introduction}
The major knowledge on the QCD phase transitions at zero baryon density comes
from first principle calculations made on supercomputers, namely from the
lattice. When simulations are run with physical quark masses, it is well
known that lattice predicts the restoration of chiral symmetry, which is
spontaneously broken by the quark condensate in the vacuum, at a finite value
of the temperature $170$~MeV $\leq T \leq 200$~MeV. The chiral restoration in
the vacuum is actually a smooth crossover, the reason being that finite
values of the quark masses break explicitly chiral symmetry, hence there is
not a true phase transition. For simplicity, from now on I will call the
chiral crossover, as well as the true phase transition, the chiral
restoration. In correspondence of the chiral restoration, lattice shows that
a deconfinement transition occurs. This has suggested that chiral restoration
and deconfinement of color are two intimately connected transitions of QCD,
(see Ref.~\cite{Bazavov:2009zn} and references therein).

Lattice investigations at finite baryon chemical potential, $\mu$, suffer the
(in)famous sign problem in three color QCD. To this end, several approximated
methods have been used to circumvent it. By means of one of these methods,
namely the two parameter reweighting, it has been predicted, some time
ago~\cite{Fodor:2004nz}, that the chiral crossover becomes a first order
transition at a certain value of $\mu$. The couple $(\mu_E,T_E)$ in the
$(\mu,T)$ plane at which this occurs is called the Critical End Point (CEP)
of the QCD phase diagram. The numerical simulations of
Ref.~\cite{Fodor:2004nz} predict $\mu_E\approx 350$~MeV and $T_E \approx 160$
MeV.

An interesting alternative to the reweighting analysis of the QCD phase
diagram, with particular reference to the existence of a CEP, has been
performed~\cite{deForcrand:2008zi} (see also references therein). The
reasoning on which the investigations of~\cite{deForcrand:2008zi} lies is
very simple to understand: at $\mu=0$, it is known, from lattice studies,
that the chiral transition is a true first order transition, if quarks are
taken in the chiral limit. Moreover, as the quark masses are increased above
a critical value, the transition becomes a crossover. It happens that at the
physical point, defined as the couple of values for the up- and strange-quark
mass, $(m_u,m_s)$, which gives the
physical spectrum of mesons, the transition is a crossover. Hence, there
exists a critical line in the $(m_u,m_s)$ plane which is the border between
an inner region, in which the chiral transition is of first order, and an
outer region, in which the transition is a crossover. As $\mu$ is increased,
one can study the evolution of the critical line in this plane. In order to
circumvent the sign problem, the authors of Ref.~\cite{deForcrand:2008zi}
performed a Taylor expansion in powers of $\mu/T$, computing all the
coefficients at $\mu=0$ (where the sign problems is absent). Within the
Taylor expansion, the critical line is expressed as
\begin{equation}
m_{c}(\mu) = m_c(0)\left[1 + \sum_{k=1}^N c_k \left(\frac{\mu}{T_c}\right)^{2k}\right]~.\label{eq:i}
\end{equation}
The coefficient $c_1$ governs the behavior of the critical line at small values of $\mu$. Nowadays, the coefficients
$c_k$ have been determined up to the $8^{th}$ order. Surprisingly enough, the results of Ref.~\cite{deForcrand:2008zi}
are that the critical line moves towards lower values (hence to less realistic) of the quark masses, as $\mu$ is
increased. This means that at finite (but small, see below) values of $\mu$ the crossover remains crossover, if quarks
are taken in the chiral limit. The analysis performed in Ref.~\cite{deForcrand:2008zi} should be reliable, by author's
admission, up to $\mu\approx 500$~MeV. As a consequence, their results are consistent with the scenery in which a CEP,
if it there exists, is located at values of $\mu$ larger than that predicted in~\cite{Fodor:2004nz}. The discrepancy is
probably due to the fact that the reweighting method suffers of large systematic errors at large $\mu$.

It is of a certain interest to compare this scenario with that of some model
calculation. Among the various models, the NJL model (or its improved
version, the Polyakov--Nambu--Jona-Lasinio~(PNJL) one) is a very
popular one (for review see~\cite{revNJL}). The NJL model Lagrangian shares
the same global symmetries of the QCD Lagrangian. Since we can describe the
numerous (expected) phases of the QCD phase diagram in terms of
broken/restored global symmetries, the hope is that the NJL calculations
grasps, for the property specified above, at least the main characters of the
QCD phase diagram in the $\mu$-$T$ plane. Moreover, determining the ground
state of the model at any temperature and/or chemical potential is a very
easy task, which requires only some numerics. On the other hand, first
principle calculations are not feasible at finite $\mu$ both numerically (for
the infamous sign problem of three color QCD) and analytically (for weak
coupling approximation might break down in the range of temperature/chemical
potential relevant for heavy ion collisions as well as for compact star
phenomenology). Therefore, the NJL model might be helpful in depicting the
main aspects of the QCD phase diagram.

The NJL (or PNJL) phase diagram has been discussed in several papers. Here I refer to~\cite{Ciminale:2007ei}. First of
all, I need to specify the model Lagrangian density,
\begin{equation}
{\cal L} = \sum_f\bar\psi_f\left(\,iD_\mu \gamma^\mu - m_f + \mu\gamma_0\right)\psi_f ~+~{\cal L}_4~+~{\cal
L}_6~,\label{eq:lagr1}
\end{equation}
where the sum is over the three flavors $f$ $(=1,2,3$ for $u,d,s$). In the above equation the background gauge field
$A_\mu = g\delta_{\mu0}A_{a\mu}T_a$ is coupled to quarks via the covariant derivative $D_\mu = \partial_\mu-i A_\mu$
and $A_\mu$ will be specified later; $m_f$ is the current mass (we assume $m_u = m_d$). The quark chemical potential is
denoted by $\mu$. The NJL four-fermion and six-fermion interaction
Lagrangians are as follows~\cite{revNJL}:
\bea {\cal L}_4 &=& G\sum_{a=0}^8\left[\left(\bar\psi \lambda_a \psi\right)^2
+ \left(i\bar\psi \gamma_5\lambda_a \psi\right)^2
\right]\label{eq:full4}~,\\
{\cal L}_6 &=&
-K\left[ \det\bar\psi_f(1+ \gamma_5)\psi_{f'} + \det\bar\psi_f(1- \gamma_5)\psi_{f'} \right]\
, \label{eq:full6}
\eea
where $\lambda_a$ are the Gell-Mann matrices in
flavor space ($\lambda_0 = \sqrt{2/3}~{\bm 1}_f$) and the determinant is in flavor space as well. The parameters are
\be
m_{u,d} =  5.5~\text{MeV}, \ \   m_{s}  =  140.7~\text{MeV},  \ \
G\Lambda^2 = 1.835,  \  K\Lambda^5 = 12.36, \ \
\Lambda = 602.3~\text{MeV}. \nonumber
\ee
From these parameters one gets $m_\pi \simeq
135$~MeV, $m_K \simeq 498$~MeV, $m_{\eta^\prime} \simeq 958$~MeV,
$m_\eta \simeq 515$~MeV and $f_\pi \simeq 92$~MeV.

Once the Lagrangian is specified, the thermodynamic potential at temperature
$T$ is obtained after integration over the fermion fields in the partition
function:\be \Omega = {\cal U}[T,\Phi,\bar\Phi]
+ \Omega_{q}[M_f,\Phi,\bar\Phi]~,
\label{eq:due} \ee where $\Omega_q$ denotes the free quark contribution, as well as the interaction term of quarks with
the Polyakov loop (see~\cite{Ciminale:2007ei} for more details). In the
thermodynamical potential, the term ${\cal U}(T,\Phi,\bar\Phi)$ is the
novelty that improves the NJL model and promotes it to the PNJL
model~\cite{Fukushima:2003fw}. It describes the dynamics of the traced
Polyakov loop in absence of dynamical quarks.  The potential ${\cal U}$
cannot be determined by first principles: one has to chose a convenient form
for it, by trying to reproduce lattice data on thermodynamical quantities of
the pure glue theory. Different analytical forms of ${\cal U}$ lead to
different quantitative predictions, even if the qualitative picture is quite
not sensible of the form chosen. In this talk I focus on a model calculation
based on the following potential: \be\frac{{\cal U}(T,\Phi,\bar\Phi)}{T^4}=
-\frac{\tilde b_2(T)}2\Bar{\Phi}\Phi\,+\,b(T)\ln[1-6\Bar{\Phi}\Phi+4(
\Phi^3+\bar\Phi^{3})-(\Bar{\Phi}\Phi)^2]\ee where the analytical form of the coefficients has been determined in
Ref.~\cite{Roessner:2006xn}.

\begin{figure}
\begin{center}
\includegraphics[width=10cm]{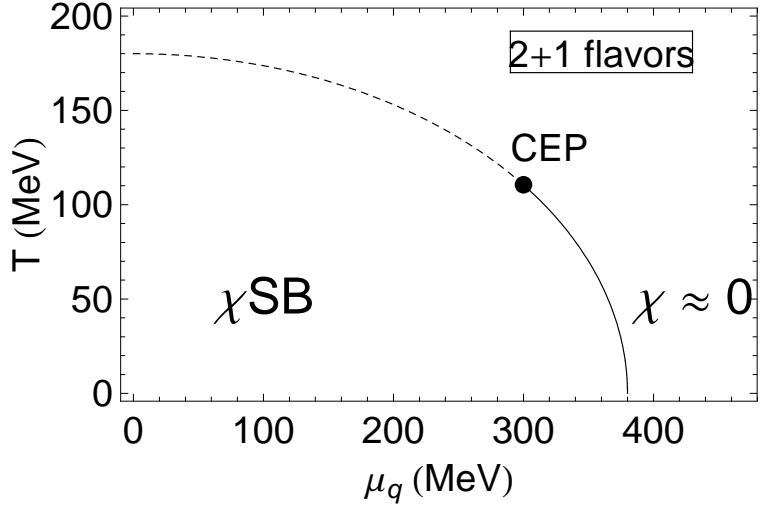}
\end{center}
\caption{\label{Fig:pd} Sketch of the phase diagram of the PNJL model with 2+1 massive flavors. Here $\mu_q$ denotes
the quark chemical potential, $\mu_q = 3 \mu$, where $\mu$ corresponds to the baryon chemical potential. For
simplicity, I have drawn only the chiral crossover line. The dashed line denotes the chiral crossover, the solid line
corresponds to a first order transition. The region denoted symbolically by $\chi_{SB}$ denotes the zone of the phase
diagram with quark condensate different from zero. In the region $\chi\approx0$, on the other hand, one has
$\langle\bar u u\rangle\approx 0$ but $\langle\bar s s \neq 0\rangle$. Based on~\cite{Ciminale:2007ei}.}
\end{figure}

 In the mean field approximation, which is formally equivalent to determine
only the classical contribution to the partition function, one can get quark
condensates $\sigma_f$ and Polyakov loop for any value of the parameters
$\mu$ and $T$ simply by looking at the global minima of $\Omega$. Depending
on the values of $\sigma_f$ and $\Phi$, one can characterize the symmetry
breaking pattern of the theory in any point of the plane $\mu$-$T$, hence one
can build a phase diagram. The phase diagram of the model is sketched in
Fig.~\ref{Fig:pd}. For simplicity, I have drawn only the chiral crossover
line. The dashed line denotes the chiral crossover, the solid line
corresponds to a first order transition. The region denoted symbolically by
$\chi_{SB}$ denotes the zone of the phase diagram with quark condensate
different from zero. In the region $\chi\approx0$, on the other hand, one has
$\langle\bar u u\rangle\approx 0$ but $\langle\bar s s \rangle \neq 0$. It is
interesting to notice that, with the parameters at hand that reproduce the
vacuum spectra of the pseudoscalar mesons, the CEP is located at quite large
values of the quark chemical potential, which is one third of the baryon
chemical potential, thus at values of $\mu$ larger than the $350$~MeV quoted
above. The introduction of a vector interaction can shift $\mu_{\rm CEP}$ to
higher values, depending on its magnitude at finite
density~\cite{Fukushima:2008wg}. It can even disappear at all, if the vector
interaction is repulsive enough. Hence, we can conclude that the PNJL model
scenario is in agreement with the newest lattice findings on the absence of a
CEP at small values of the baryon chemical potential.

%\section{Conclusions}

\acknowledgments I acknowledge H. Abuki, M. Ciminale, P. Colangelo, R. Gatto, M. Mannarelli and S. Nicotri for valuable
collaboration and interesting discussions on the topics discussed here. The original publication is available at
``http://www.sif.it/SIF/en/portal/journals''. This talk is dedicated to the memory of my mentor, prof. Giuseppe
Nardulli.

\end{document}